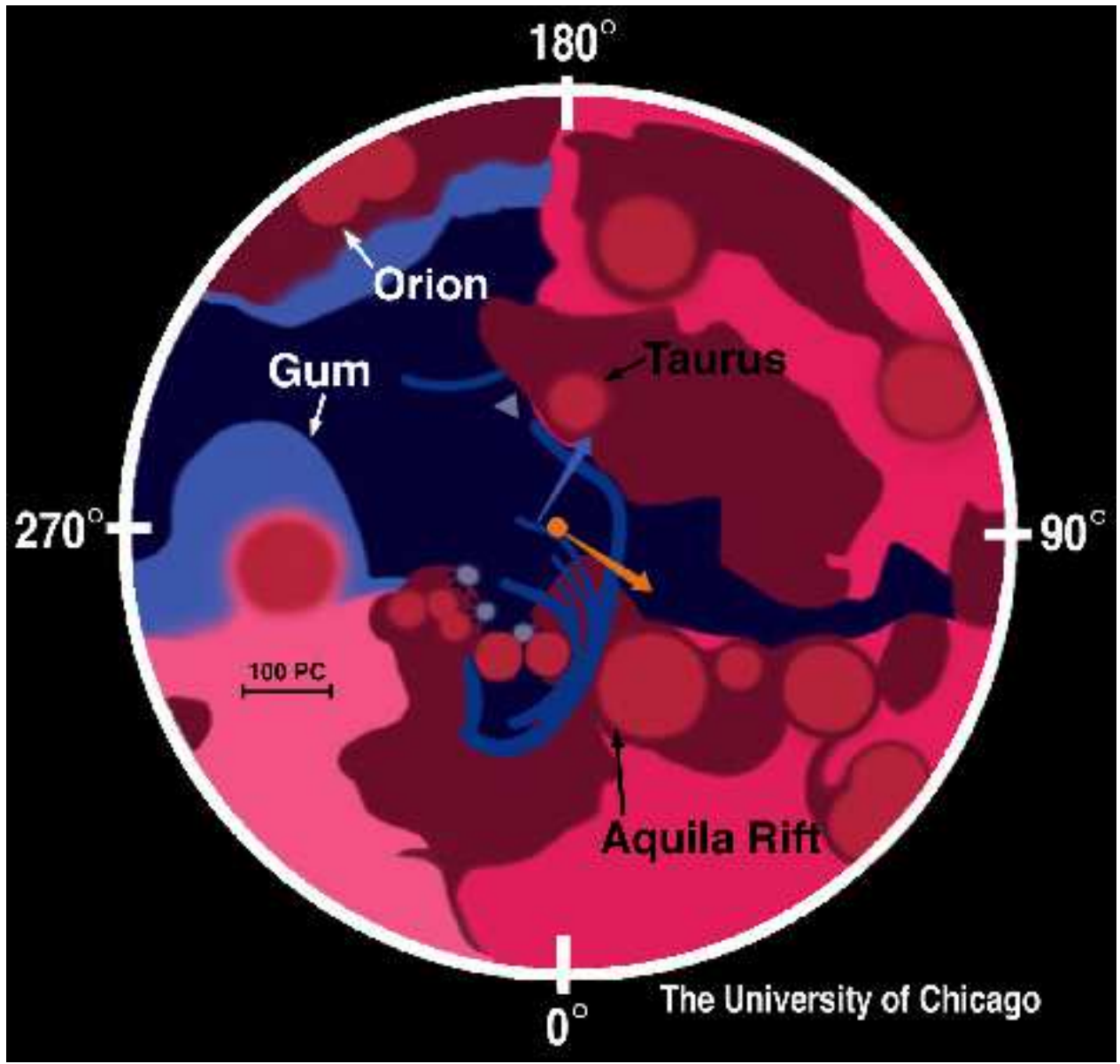

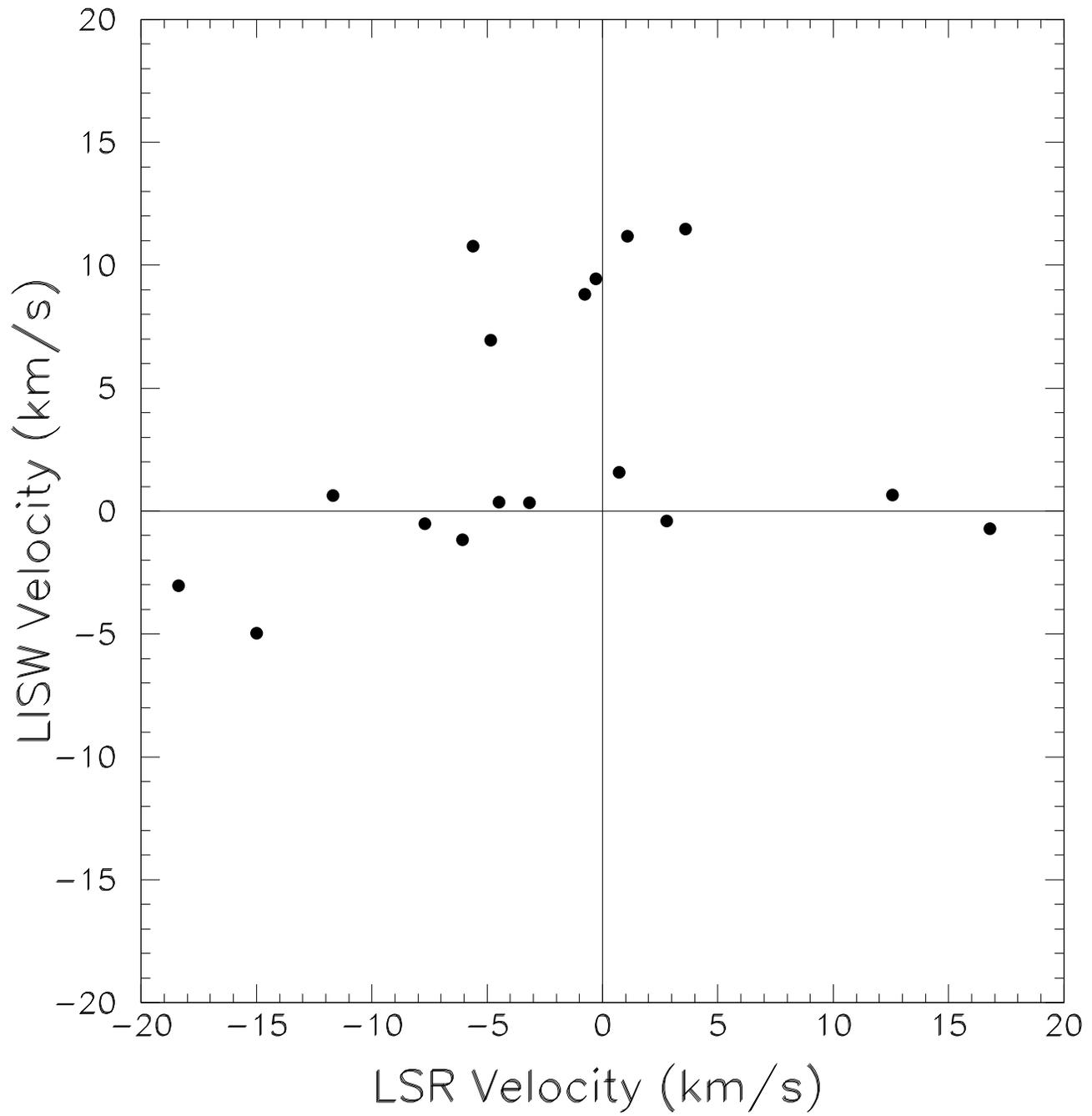

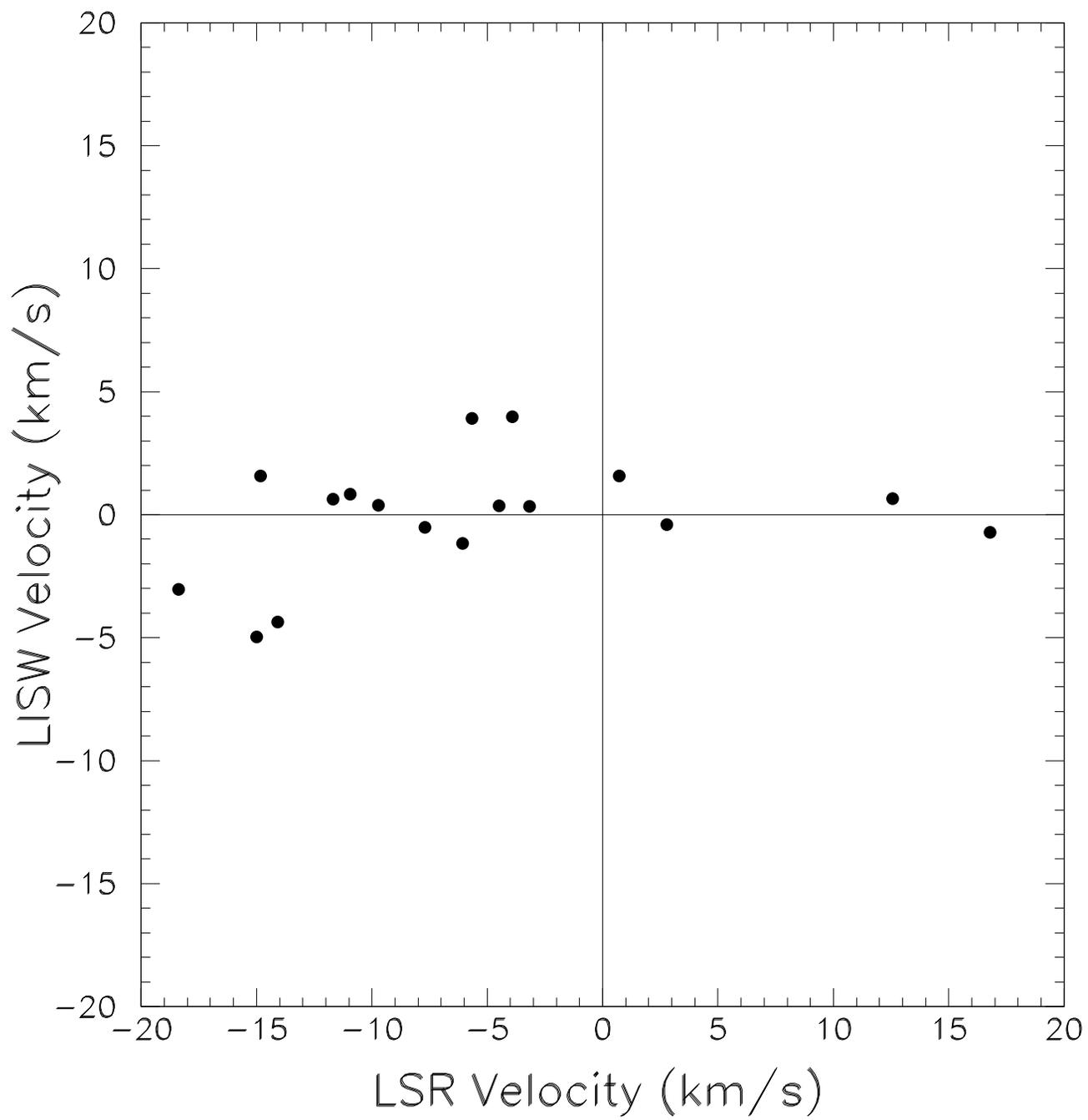

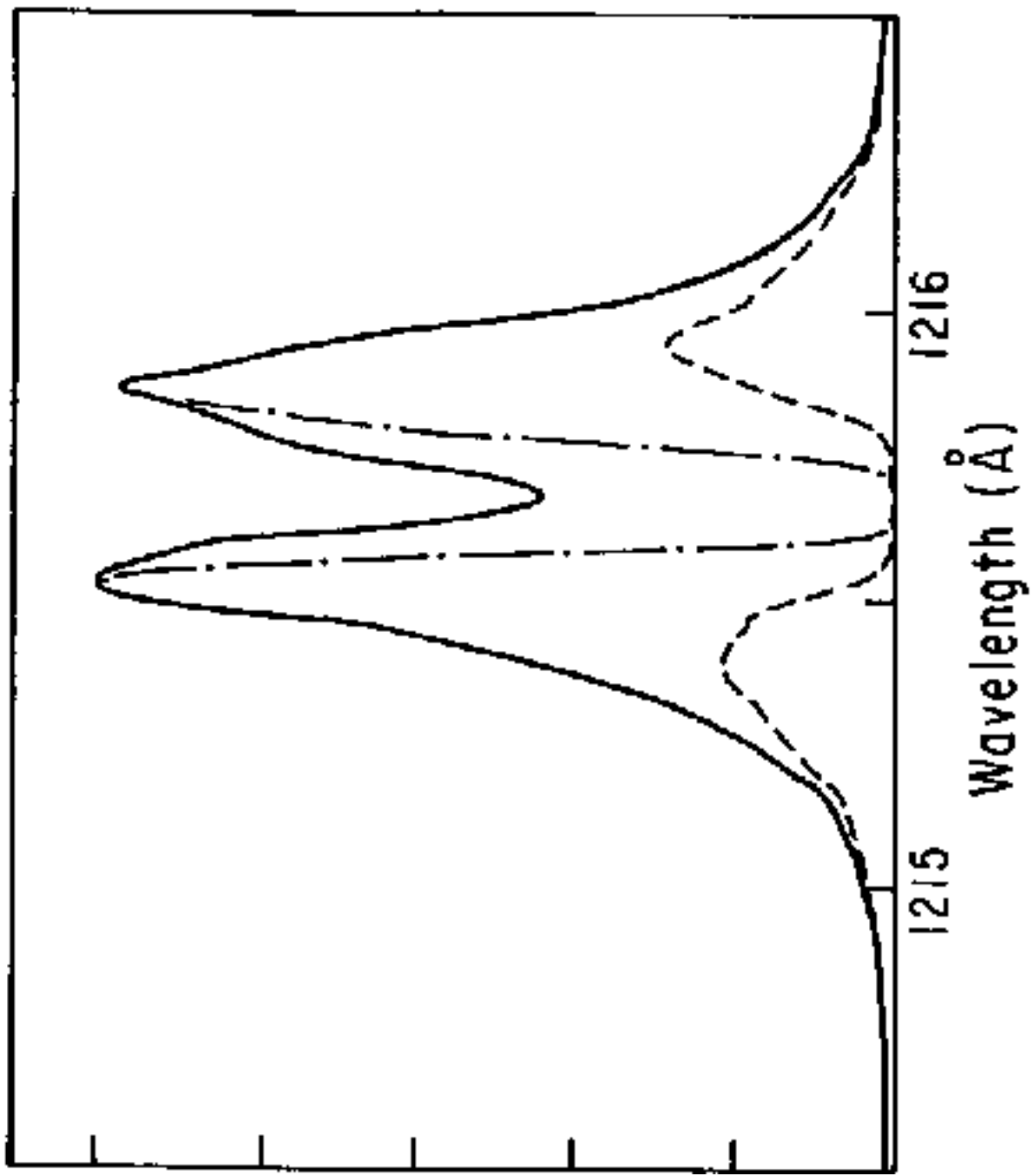

Wavelength (Å)

# The Journey of the Sun

Priscilla C. Frisch
Department of Astronomy and Astrophysics
The University of Chicago
5640 South Ellis Avenue, Chicago, IL   60637
Phone:   773/702-0181;   FAX:   773/702-8212
email:   frisch@oddjob.uchicago.edu
Preprint Series:   March 17, 1997;   Revised:   May 20, 1997
Enrico Fermi Institute Preprint No. 97-23

## Abstract

An analysis of the distribution and kinematics of interstellar material within 500 pc of the Sun leads to the conclusion that the galactic environment of the Sun changes with time.   Consideration of evidence for interstellar gas interacting with the solar wind implies that these variations may alter the interplanetary environment of the Earth. The events causing the $^{10}$Be spikes 33,000 years and 60,000 years ago in the Antarctic ice record must be associated with the Local Fluff cloud complex, possibly due to solar encounters with structures with subparsec scale sizes. An encounter with "dense" interstellar cloud material could attenuate the solar L$\alpha$ flux by as much as 70%, modify mesospheric chemistry, modify the magnetosphere-solar wind coupling, and alter the global electrical circuit.   Prior to the entry of the Sun into the Local Fluff cloud complex, within the past 200,000 years, the galactic environment of the Sun differed radically from the environment prevailing today and expected for the near future.

# Introduction

The galactic environment of the Sun influences the one astronomical unit (AU) interplanetary environment around the Earth, including the Sun-Earth coupling mechanisms and the high energy radiation and particle fluxes at the top of the Earth's atmosphere. Our improved understanding of the interaction between the solar wind and heliosphere with the surrounding interstellar cloud, including the invasion of interstellar gas into the inner heliosphere via the daughter pick-up ion and anomalous cosmic ray products, suggests that it is time to take a new look at the galactic environment of the Sun on its journey through space.

During the past five million years the Sun has been immersed in a region of space, with very low interstellar densities, between the Orion spiral arm and the spiral arm spur known as the "Local Arm". [1] The interarm region is bounded by molecular clouds, diffuse gas, and superbubbles expanding from stellar nurseries in adjacent spiral arm molecular cloud material. This low density region coincides with the interior of the band of young stars known as Gould's Belt. In this paper I refer to this low density region as an "interarm" region. It is interesting that the part of the celestial sky viewed from the northern terrestrial hemisphere contains more interstellar material than does the portion viewed from the southern hemisphere, since the North Celestial Pole is oriented towards galactic coordinates l=123°, b=+27°. In the northern hemisphere we see interstellar matter associated with the trailing edges of the Perseus and Orion spiral arms, whereas the interarm region is viewed from the southern hemisphere.

Our galactic environment changes with time since the Sun moves through space with a velocity of ~16.5 pc per million years. Observations of AU sized "dense" ($>10^5$ $cm^{-3}$) structures in interstellar gas show environment variations are possible on time scales of 3 years to 3,000 years, and on even shorter time scales when effects from supernova radiation are included.

The interaction between the solar wind and interstellar matter regulates the interplanetary environments within the solar system. Today interstellar gas flows through the solar system at a rate of about ~5 AU per year. The solar wind prevents interstellar gas from penetrating to the inner heliosphere, thereby shielding the Earth. At 1 AU the solar wind has a density of ~5 particles $cm^{-3}$, whereas the inflowing neutral interstellar hydrogen (density ~0.15 $cm^{-3}$) is excluded from the inner heliosphere because it becomes fully ionized ~3-5 AU from the Sun. In contrast, the interplanetary environments of the outer planets rapidly (on timescales of decades) respond to changes in interstellar gas densities. The heliosphere was certainly larger several million years ago when the Sun was in the interarm region, and will shrink if a denser, more highly ionized, or faster cloud were encountered. Happily observations of the outer planets and *in situ* observations of pick-up ions anomalous cosmic rays and neutrals in the outer heliosphere would provide advance notice of such an occurrence.

In this paper the distribution and properties of nearby interstellar clouds are examined to assess the galactic environment of the Sun over time scales of ±5 Myrs. Cloud distribution and kinematics provide the basis for evaluating encounters between the Sun and interstellar clouds. Previous studies have linked specific galactic events to anomalies in the geologic record of the Earth, including attributing organic layers deposited ~11,500 years ago to the influence of gamma-rays from the Vela supernova explosion on mesosphere chemistry[2], and attributing spikes in $^{10}$Be concentration 33 kyrs and 60 kyrs ago in Antarctic ice cores to enhanced cosmic ray fluxes at the top of the Earth's atmosphere[3]. Additional mechanisms for an influence of interstellar gas on terrestrial climate are suggested, and possible astrophysical explanations for the $^{10}$Be spikes are discussed.

## Space Motion of Sun

A discussion of variations in the solar galactic environment must be based on comparisons between the solar space motion and the distribution and kinematics of interstellar clouds. The aspect of the solar motion relevant here is the motion of the Sun with respect to nearby interstellar clouds, which on the average are at rest in the local standard of rest (LSR).[4] For the motion of the Sun through the LSR, I will use ~16.5 km s$^{-1}$ towards the direction l=53$^{o}$, b=+25$^{o}$, corresponding to a motion of 16.5 pc per million years through the LSR. This value is based on observations of the velocity ellipsoid of nearby cool stars which have relaxed to dynamical equilibrium in the galactic gravitational field.[5] For comparison, the frequently used Standard Solar Motion is 19.5 km s$^{-1}$ towards l=56$^{o}$, b=+23$^{o}$, which is based on commonly observed bright stars including bright B stars which have a residual peculiar radial velocity ~5 km s$^{-1}$ due to Gould's Belt kinematics[6]. The discussions in this paper are relatively insensitive to the difference between these two motions, which introduces uncertainties in the solar position of ±15 pc over a period of ±5 Myrs. An additional uncertainty due to a possible LSR radial motion of 14 km s$^{-1}$ outflowing from a bar structure in the inner galaxy is not relevant since the relative space trajectories of the Sun and clouds would not change. The 66 Myr period of oscillation of the Sun around the galactic plane and the 250 Myr orbit of the solar neighborhood about the galactic center are also not significant here.

### Global Distribution and Kinematics of Nearby Molecular Gas

To evaluate possible encounters between the Sun and interstellar clouds over time scales of ±5 Myrs, the global distribution and kinematics of nearby interstellar material must be determined. Molecular clouds define the positions of spiral arms, and globally ~50% of the mass of interstellar matter is in molecular clouds. Fig. 1 shows the positions of molecular clouds using the cloud list in Dame et al.[7] based on observations of the 115 GHz 1–>0 rotational CO transition. The molecule CO is a standard tracer for clouds with densities $10^3$-$10^6$ cm$^{-3}$. The viewpoint is the North Galactic Pole, looking "down" onto the galactic plane, except that interstellar matter towards Orion is ~15$^o$ below the plane, while towards Scorpius it is ~15$^o$ above the plane, consistent with the local 10$^o$-20$^o$ warp of the galactic plane associated with Gould's Belt.

The molecular clouds shown in Fig. 1 are basically at rest in the LSR. The mass-weighted average LSR velocity of these clouds is 2.9 ±0.6 km s$^{-1}$, where 0.6 km s$^{-1}$ is the standard deviation of the mass-weighted molecular cloud velocities. This average is calculated from radial velocities and masses of CO clouds within 500 pc (Table 2, Dame et al.), for the clouds Aquila Rift, Cloud B, Vulcan Rift, Lindblad Ring, Cepheus, Taurus, Perseus, Orion A, Orion B, Vela Sheet, Chameleon, Coal Sack, G317-4, Lupus, ρ Oph and RCrA. The mass-weighted average velocity is $\Sigma \mid (M_i\ v_i)\mid / \Sigma\ M_i$, where the summation is over clouds and $M_i$ and $v_i$ represent the mass and velocity of the selected clouds, respectively. If the average density of these clouds is $10^3$ cm$^{-3}$, the "filling factor" (fraction of space filled by cloud type) of CO molecular clouds within 500 pc of the Sun and 50 pc of the galactic plane is 0.16.

The deficiency of interstellar matter in the third Galactic quadrant between l=220$^o$ and l=250$^o$ corresponds to the interarm region between the Local and Orion Arms, and merges with the interarm region between the "grand design" Orion and Sagittarius spiral arms at about 1 kpc from the Sun. This deficiency of interstellar matter is shown by a lack of 21 cm emission[8], diffuse gamma-ray emission[9], molecular gas, and diffuse gas and dust[10,11]. For the past few million years this interarm region constituted the galactic environment of the Sun.

The Gum nebula[12] is shown in Fig. 1. The Vela Sheet (l=272$^o$-279$^o$, b=−3$^o$) is embedded in the Gum nebula, a region of ionized filaments and clouds centered near l=271.3°, b=+5.3°, and with a diameter of ~36°. An abrupt increase in reddening at 200 pc is seen in directions l=258.4°, b=−11.1° l=271.3°, b=+5.3°, which is attributed to the onset of reddening due to dust in the Gum nebula.[13] If spherical, the Gum extends from 200 to 380 pc from the Sun. The star ε CMa (see below) may be embedded in the edge of the Gum nebula, indicating that except for the Local Fluff complex, the Gum has little foreground H° (see below). Assuming that the Vela Sheet is centered in the Gum Nebula, it is plotted at a distance of 200 pc in Fig. 1.

### Global Distribution and Kinematics of Nearby Diffuse Gas

Interstellar gas and classic dust grains (radii ~0.1-0.3 μm) are closely correlated in space, so observations of the reddening of starlight by interstellar dust trace the distribution of diffuse gas of densities 1-1000 cm$^{-3}$ (N(H)/E(B−V)=5.8 x 10$^{21}$ cm$^{-2}$ mag$^{-1}$ [14]). Fig. 1 shows the distribution of diffuse dust and gas, based on the color excess reddening index E(B−V)[10]. The minimum contour E(B−V)=0.1 mag, corresponds to a hydrogen column density 5.8 x 10$^{20}$ cm$^{-2}$. As expected, diffuse gas is associated with spiral arm molecular material.

The first three parsecs in the anti-apex direction contain interstellar gas which is part of the Local Fluff complex of clouds surrounding the Sun. Beyond that, space is essentially empty. Towards the distant CMa stars, n(H°)=0.00004 cm$^{-3}$.[15] The stars α CMa (d=2.7 pc, l=227°, b=−9°) and ε CMa (d=187 pc, l=240°, b=−11°) are both within 17° of the anti-apex direction (l=236°, b=−23°). These two stars have similar total H° column densities. Towards α CMa N(H°)=3.2 10$^{17}$ cm$^{-2}$, and towards ε CMa N(H°)=3.0 10$^{17}$ cm$^{-2}$, indicating that most of the H° is within several pc of the Sun[15,16]. Ninety-two percent of the H° towards ε CMa is associated with the two cloud components seen in front of α CMa, and outside of the Local Fluff cloud complex n(H°)<4 x 10$^{-5}$ cm$^{-3}$ towards ε CMa. Similar low H° column densities are seen towards the white dwarf star pair RE0457-281 and RE0503-289 (with l=229°, b=−36°, d=90 pc, N(H°)=7.2 x 10$^{17}$ cm$^{-2}$) and β CMa (with l=226°, b=−14°, d=220 pc, N(H°)=5-9 x 10$^{17}$ cm$^{-2}$).[25] Extrapolating backwards in time, at a space velocity of ~17 pc per Myrs, the Sun would have encountered the Local Fluff cloud complex within the past 200,000 years.

The Local Bubble is also characterized by low electron densities ($n(e^-) < 0.02$ cm$^{-3}$ [17]) with larger electron densities outside of the Local Bubble region[18].

The Na$^o$ and Ca$^+$ optical absorption lines provide an easily observed tracer of diffuse clouds, typically sampling clouds with densities 0.1 cm$^{-3}$ to 100 cm$^{-3}$, and thermal temperatures 50$^o$ to 8,000$^o$. For over 80 absorption line components observed in stars within 500 pc, Vallerga et al.[19] found mean and RMS LSR velocities for absorption components of 0.9 ±11.3 km s$^{-1}$ for line components seen only in Ca$^+$, 0.8 ± 4.7 km s$^{-1}$ for components seen both in Na$^o$ and Ca$^+$, and 0.6 ± 3.6 km s$^{-1}$ for components seen only in Na$^o$. For all three line-sets the mean velocity is approximately 0 km s$^{-1}$. A similar result is found by the statistical analysis of an ensemble of northern hemisphere neutral hydrogen clouds observed in the H$^o$ 21 cm line, where the clouds are found to be at rest in the LSR with a velocity dispersion of 6.9 km s$^{-1}$.[20]

A critical issue affecting solar encounters with dense interstellar clouds during the next 50,000-10$^6$ years is whether the cloud complex flowing past the solar system is at rest in the LSR. Systematic flows in nearby interstellar gas have been inferred both over 100 pc scale sizes[21,22] and 30 pc scale sizes[19,23,24,25]. The local flow is easily seen when the velocities of Ca$^+$ absorption components formed within 30 pc of the Sun are compared in the LSR versus local flow velocity frames. The velocities of Ca$^+$ components in 17 nearby stars[26] are plotted in the LSR versus the local interstellar flow velocity frames in Fig. 2 using Ca$^+$ observations[19,23,24,27,28,29]. These 17 stars show 36 absorption components (at spectral resolution 0.5-3 km s$^{-1}$), but only one component per star is plotted. In Fig. 2a the plotted components are selected to be the components with the smallest velocity in the LSR, and the components have average velocities $-5.5 \pm 9.8$ km s$^{-1}$ in the LSR. In Fig. 2b the plotted components are selected to be those which have the smallest velocities in the interstellar flow frame, where they have average velocity $-0.1 \pm 2.2$ km s$^{-1}$ in the flow frame (1 sigma deviation given). The heliocentric flow vector $-26.8$ km s$^{-1}$ from the upwind direction l=6.2$^o$, b=+11.7$^o$ is used in Fig. 2b.[24] This vector represents an all-sky average velocity for the nearest stars, while ignoring possible small scale velocity gradients such as identified by the LIC (for "local interstellar cloud") versus G (for "galactic") vectors of Lallement and Bertin. The data were acquired at spectral resolutions 1.5 km s$^{-1}$ to 3 km s$^{-1}$. Comparing the average velocities and dispersions of the selected components in the local flow versus LSR velocity frames, it is clear that the local flow velocity frame is a much better fit to $Ca^+$ absorption components in nearby stars than is the LSR velocity frame, although the fit is not perfect.

The distribution of velocity components about the flow velocity flags cloud-cloud velocity differences. The dispersion $\pm 2.24$ km s$^{-1}$ about the flow velocity indicates that intra-cloud velocity differences within the local cloud complex are mainly subsonic (at T~7,000 K, the sound velocity is 10 km s$^{-1}$). In the cloud surrounding the solar system, $n(e^-)$~0.15 cm$^{-3}$, giving an Alfven velocity of 8.4 km s$^{-1}$ for B~1.5 µG. Among the 36 components considered, 25% have velocities in the flow frame larger than 8 km s$^{-1}$. If $n(e^-)$~0.15 cm$^{-3}$ were uniform in the Local Fluff cloud complex, most cloud-cloud interactions would therefore also be sub-Alfvenic. However, the second (blue-shifted) cloud seen towards α CMa and ε CMa has $n(e^-)$~0.46 cm$^{-3}$ and T=3,800 K, giving sound and Alfven velocities of 7 km s$^{-1}$ and 5 km s$^{-1}$, respectively (for assumed magnetic field strength ~1.5 µG[30]). Supersonic cloud-cloud collisions are therefore possible in the local cloud complex.

### Superbubble Shells

Fig. 1 shows the location of shells associated with star formation in Orion and Scorpius-Centaurus. The combined mechanical energy input from stellar winds and supernova from massive stars in star forming regions create "bubbles" consisting of shell-like diffuse clouds surrounding relatively empty regions of space. Stellar associations evolve on timescales of ~5 Myrs, so the diffuse gas distribution may change significantly on longer time scales. It is a characteristic of the expansion of these superbubble shells that they have penetrated to greater distances into the interarm regions than in directions requiring displacements of large masses of interstellar clouds. The radius of a superbubble shell scales as $n_0^{-0.2}$, where $n_0$ is the preshock density, so that a shell expanding into an interarm region (n~0.0004 cm$^{-3}$) will expand a factor of 30 further than a shell expanding into a molecular cloud (n~10$^3$ cm$^{-3}$). This property may reconcile the asymmetric distribution of local interstellar matter with the symmetry of the soft X-ray background. It has been suggested that the Sun is embedded in one of the superbubble shells associated with the formation of the Scorpius-Centaurus Association.[31]

"Orion's Cloak" is a fast ($\sim -100$ km s$^{-1}$) radiative shock 300-350 pc from the Sun in front of the Orion-Eridanus region (l~200$^\circ$ to l~215$^\circ$), and propagating into very low density gas. The nearside of this bubble is placed at 140-200 pc from the Sun.[32,33,34]   The formation of Orion's Cloak may be associated with the explosion which formed the Geminga pulsar, which is runaway from a supernova explosion in the Orion association.[35,36]   The position of the ~300,000 year old Geminga pulsar is shown in Fig. 1 at 157 pc (+59, -34).[37]   At this distance, and with an origin in the Orion Association, Geminga is moving at ~1,000 km s$^{-1}$ and will reach the solar vicinity in about 150,000 years.

Superbubble shells expanding into an inhomogeneous medium due to three successive epochs of star formation in the Scorpius-Centaurus Association (SCA) are shown in Fig. 1.   The star formation epochs correspond to  the formation of the Upper Centaurus Lupus subgroup (14-15 Myrs ago), the Lower Centaurus Crux subgroup (11-12 Myrs ago) and the Upper Scorpius subgroup (5-6 Myrs ago).[38] The largest of these shells is the oldest, and the interior of this shell corresponds to the region commonly denoted the "Local Bubble".   The large scale H$^\circ$ filaments[39] seen in the galactic center hemisphere of the sky are modeled as superbubble shell fragments from the successive epochs of star formation in the SCA expanding into an inhomogeneous medium.[25]   The shell fragments from these three star-formation epochs are illustrated in Fig. 1.   Shells from the first two star-formation epochs have expanded past the solar location. The 4 Myr old shell from the third star-formation epoch has expanded into the evaporated gas from residual molecular clouds embedded in the association, and now corresponds to the upwind gas sweeping down on the solar location.   The Sun is displaced above (with respect to galactic coordinates) most of the mass in the 4 Myr shell within 10 pc.   Shell geometries are based on observations of H$^\circ$ filamentary structures at positive galactic latitudes in the galactic center hemisphere of the sky, and negative galactic latitudes in the anti-center hemisphere of the sky.

### Sun-Cloud  Encounters

The distribution and kinematics of nearby interstellar clouds can now be compared with the path of the Sun through space.   A solar velocity of 16.5-19.5 km s$^{-1}$, corresponds to a space motion of 16.5-19.5 pc per million years, so timescales of $\pm 10^6$, $10^5$, and $10^4$ years correspond to distances of ~20, 2 and 0.2 pc, respectively.   From Fig. 1 we see that if the interstellar cloud distribution were static in time, the Sun would not have encountered any neutral interstellar gas for 10-20 million years, except for the Local Fluff cloud complex. Two regions of ionized gas at unknown distances, each ~6.4 pc thick, are found towards $\beta$ CMa, within 15 pc of the anti-apex direction.[40] Since these clouds are not observed in $\epsilon$ CMa, they may be local to $\beta$ CMa, but the distance is unknown. It is not known if the Sun passed through either of these features.

About 50% of the interstellar gas of our galaxy is in molecular clouds. However, these clouds fill only ~16% of nearby space, which combined with the low velocity dispersion of nearby dense CO molecular clouds indicate that it is unlikely that the path of the Sun has intersected a molecular cloud within the past 5 million years. The three-dimensional space velocities of the individual molecular clouds in Fig. 1 are not known since only the Doppler shift of the radial component of the velocity vector can be measured. The radial velocities range from −6 to 10 km s$^{-1}$. If the cloud velocities perpendicular to the line of sight are equal to the radial velocity, then 97% of the clouds will have space velocities of less than 6 km s$^{-1}$, corresponding to motions less than 30 pc per 5 million years.

It is more likely that the Sun will encounter fast clouds, such as the warm kinematically disrupted interstellar clouds characterized by relatively large $Ca^+/Na^o$ ratios, than slow clouds. Using the velocity dispersion found by Vallerga et al. for nearby clouds with high $N(Ca^+)/N(Na^o)$ ratios (0.9 ±11.3 km s$^{-1}$), a cloud with a 2-sigma velocity 22.6 km s$^{-1}$ would have traveled about 100 pc through space over the past 5 million years, while a cloud observed only in $Na^o$ will typically have traveled less than 35 pc over the same period. The cloud complex surrounding the solar system today is characterized by high abundances of refractory elements.[31,41]

Within the past 1-2 x 10$^5$ years the Sun appears to have entered the Local Fluff cloud complex, which appears to be part of the superbubble shell outflow from the Sco-Cen Association. With the exception of the cloud surrounding the solar system, the three-dimensional space trajectories of nearby diffuse clouds are not known. Two interstellar clouds are seen in the anti-apex direction towards $\alpha$ Cen and $\epsilon$ Cen, the local cloud at 17-19 km s$^{-1}$ and a blue shifted cloud at 10-14 km s$^{-1}$.[15,16] The morphology and three-dimensional velocity of the blue-shifted cloud are not known, but it has a larger neutral column density towards $\alpha$ Cen ($N(H^o)$=1.7 x 10$^{17}$ cm$^{-2}$) than towards $\epsilon$ CMa ($N(H^o)$=7.6 x 10$^{16}$ cm$^{-2}$). For density $n(H^o) > 0.1$ cm$^{-3}$ the cloud is thin in the sightline, less than 0.25 pc. It seems likely the Sun passed through the blue-shifted cloud given that the cloud extends over13° in latitude. Using the average column density $N(H^o) \sim 5.5 \times 10^{17}$ cm$^{-2}$ towards the stars α CMa, β CMa, ε CMa, RE0457-281, and RE0503-289,[25] which are within 20° of the anti-apex direction (l=233°, b=−25°) from which the Sun is moving, and assuming an average space density of $n(H^o) = 0.1$ cm$^{-3}$, gives the estimate that if the gas were at rest in the LSR, the Sun would have first entered the local cloud complex sometime within the past100,000-200,000 years. The star β CMa has the highest neutral column density of these 5 stars, giving an encounter epoch with the Local Fluff complex of180,000 years ago.

When did the Sun encounter the interstellar cloud surrounding the solar system? The three-dimensional space motion of the cloud surrounding the solar system is known well. However, the morphology of the cloud is unknown. When the solar apex direction (l~53°, b~25°) is compared to the LSR upwind direction for the local flow (l~332°, b~+5° [25]) it is seen that the Sun and LIC cloud motions are nearly perpendicular in space. There are two consequences of this. First, nearby interstellar gas in the galactic center hemisphere of the sky sweeps past the solar location. Second, the epoch of encounter between the Sun and LIC may be highly sensitive to uncertainties in the directions of these two motions.

The simplest model for inferring the date of encounter between the Sun and LIC cloud is to assume only the cloud density ($n(H^o)$=0.2 cm$^{-2}$), the relative Sun-cloud velocity (26 km s$^{-1}$), and the column density towards the cloud surface of $N(H^o)$=1.7-2.2 x $10^{17}$ cm$^{-2}$ (α CMa and ε CMa data). In this case, the Sun would have entered this cloud 10,000-14,000 years ago, and the LIC cloud surface towards α CMa is ~0.27 pc away today. This motion is tantamount to assuming that the cloud moves in a direction parallel to a normal to the cloud surface in the rest frame of the Sun.

However, the rest frame of the Sun is not the preferred frame of reference for describing interstellar cloud motion. Instead, the LIC cloud morphology has been inferred by assuming the cloud velocity vector is parallel to a normal to the cloud surface in the LSR, such as would be expected if the surrounding cloud is a fragment of a superbubble shell expanding from the Scorpius-Centaurus Association.[42] Updating this discussion using recent results on the LIC density, $n(H^o) \sim 0.2$ cm$^{-2}$, the $H^o$ column density towards α CMa, and LSR velocity 19 km s$^{-1}$ from l=332° (the latitude, +5°, is not relevant here), gives an encounter date 1,300-2,100 years ago. The uncertainty encompasses the LIC velocities found from Ulysses He$^o$ observations[43] and optical absorption line data. If the Standard solar motion had been used instead, these times would be shortened by ~20%. The largest uncertainty is the unknown geometry of the cloud, and for the expanding shell geometry to be correct the Local Fluff has to have complex small scale structure. It is encouraging that the magnetic field between l=270° and l=0°, seen in polarization measurements of nearby stars[44], has a direction ~70°, which is nearly parallel to the cloud surface found from the expanding shell assumption.

What is the possibility that the Sun will encounter a relatively dense cloud within $10^5$-$10^6$ years? Globally, about 15% of the cold diffuse interstellar gas is in small (<100 au) dense (>$10^3$ - $10^5$ cm$^{-3}$) clouds[454] The Local Fluff cloud complex is highly structured[46], allowing the possibility that a dense cloud fragment is contained in this cloud system sweeping past the solar position. If cool cloud abundance patterns prevail (N(Ca)/N(H)=1.3 x $10^{-9}$), this fragment could have a column density of up to $10^{19}$ cm$^{-2}$, corresponding to a Ca$^+$ K line equivalent width of 1.5 mA, and not violate observations of Ca$^+$ in nearby stars. This would correspond to a 5 AU thick n~$10^5$ cm$^{-3}$ cloud. Ultraviolet observations of stars in the upwind direction are not adequate to rule out the presence of such a small dense cloud fragment, but a large dense cloud is ruled out by observations. Locally ~1 velocity component per 1-2 pc is seen in observations of α Aql and α CMa at instrumental resolution ~3 km s$^{-1}$.[16,47] The complex of clouds sweeping past the solar system extends ~30 pc towards the galactic center hemisphere (and only a few parsecs in the anti-center hemisphere), so over the next ~$10^5$ years ~2 pc of the upwind gas will sweep over the solar position. The Sun appears to be within 10,000 AU of the LIC cloud boundary in the direction of α Cen,[48] a conclusion consistent with detailed fits to the Lα profile of this star[49], and this cloud boundary should therefore pass the solar position in 2,500 years.

During passage through the interarm region, the size of the heliosphere is likely to have been at least a factor of ten larger than at present, given similar magnetic field strengths, because electron and neutral densities in the confining gas were at least ten times smaller. The density of the solar wind falls off as radius squared, and near Jupiter's orbit solar wind and interstellar densities are equal, so that the environments of outer planets will be modified by increases in interstellar densities before the environments of the inner planets.

## Cosmic Ray Increases 33 Kyr and 60 Kyr Ago

Sonett et al.[3] have interpreted spikes in the $^{10}Be$ record found in Antarctic ice cores at ages corresponding to 33 and 60 Kyr ago (the "D1" and "D2" events respectively) as indicating the passage of cosmic-ray laden interstellar shock fronts across the heliosphere. The result was that the cosmic ray flux at the top of the Earth's atmosphere was increased. Utilizing the width and shape of the $^{10}Be$ concentration as a function of time, they inferred a shock front thickness (between 1/e intensities) of 2-8 kyr for the D1 event.

Exploring the astrophysical basis for such a hypothesis, one can first conclude that for a shock velocity of 100-600 km s$^{-1}$, the shock would be 3-12 pc from the Sun today and therefore associated with the Local Fluff cloud complex. The spectral signature of this shock should be seen in some nearby stars. The only very high velocity feature seen in nearby stars of which the author is aware is the detection of a transient ~3 mA Ca$^+$ (K) line at ~85 km s$^{-1}$ (local standard of rest velocity) towards the nearby star HR 3131 (l=237$^{\circ}$, b=+6$^{\circ}$, d=43 pc). Subsequent reobservation of Ca$^+$ in this star failed to detect this feature[50], although transient Ca$^+$ features have been seen to be associated with interstellar shocks elsewhere. If real, this feature may correspond to the cooling gas behind a radiative shock. The 85 km s$^{-1}$ velocity, combined with the 2-8 kyr width of the D1 event, would indicate a width for the D1 shock of 0.17 - 0.7 pc. For a nominal ratio N(Ca$^+$)/N(H$^{\circ}$)~10$^{-8}$, such as found typically for gas in the Local Fluff cloud complex, the radiative Ca$^+$ region would have a column density of ~$3 \times 10^{18}$ cm$^{-2}$, and a space density of 1.4-5.5 cm$^{-3}$. The difficulty with this interpretation is that the passage of a 85 km s$^{-1}$ shock front across the Local Fluff cloud complex very recently should have significantly disrupted nearby interstellar gas kinematically, so that the observed low density low velocity should not be present.

A second simple interpretation of the $^{10}Be$ spike would be that the cosmic rays are magnetically confined by a subsonic structure comoving with the interstellar gas surrounding the solar system. In this case, simple estimates can be made about the size of this structure. The structure motion would thus be ~20 km s$^{-1}$, so that a 2-8 kyr width corresponds to features of width ~0.04-0.16 pc. This would mean that the gyroradius of the cosmic rays would have to be less than 0.04-0.16 pc. For cosmic ray protons, the gyroradius is r(pc)~9.55 10$^{-5}$ p(GeV)/B($\mu$G). (p(GeV)= energy in units GeV;  B($\mu$G) =

magnetic field strength in units μG). Therefore, for B = 2 μG, cosmic rays with energies of less than 3 TeV would be confined. An increase in cosmic ray fluxes for energies 3 GeV-3 TeV should be felt at the Earth since the solar-cycle correlated Forbush decrease in cosmic-ray fluxes at the Earth does not affect cosmic rays with energies greater than 3 GeV.

If the structure is comoving with the cloud surrounding the solar system, the structure associated with the D1 and D2 events, respectively, would now be 0.67 and 1.2 pc from the location of the Sun at the time of structure passage. The Sun itself moves ~0.2 pc per thousand years. In a model for the cloud surrounding the solar system, where the cloud motion is assumed to be moving in a direction parallel to a normal to the cloud surface, the Sun is found be very close to the cloud surface in both the upwind and downwind directions, so that the Sun is in a filament of gas less than a parsec thick with a motion in a direction parallel to the elongation of the filament. This model may not be correct, but it is consistent with the narrow spatial structures inferred by the $^{10}$Be widths.

The $^{10}$Be spikes may be also due to an increase in neutral interstellar densities, rather than galactic cosmic ray fluxes. If the interstellar neutral density in these structures were significantly larger than now, which is possible but not necessary, it is likely both that the size of the heliosphere would have decreased (roughly as the square root of the relevant interstellar ram, plasma, and cosmic ray pressure terms), and that the inflow of interstellar neutrals would have increased. Thus the density of anomalous cosmic rays at the Earth's orbit should also have increased. This interpretation offers a natural explanation as to why no $^{10}$Be spikes are seen in the ice core record prior to 60 kyrs ago, as the Sun would have been in the third-quadrant void at that time.

### Possible Effects on Terrestrial Climate

The most interesting question is whether an encounter with a relatively dense cloud may perturb the terrestrial climate. Although the detailed behavior of the terrestrial climate equilibrium is not well understood, general statements are useful. The solar wind effectively shields the inner heliosphere and the 1 AU interplanetary environment of the Earth from possible climatic perturbations due to the LIC cloud. However, with passage through a dense cloud the interaction between the solar wind and interstellar gas would be modified. Previous discussions include suggestions that an encounter with a cloud of n~$10^3$ cm$^{-3}$, B~5 μG, 80 K, could collapse the heliosphere to within ~1 AU (depending on relative velocities)[51,52], and that the explosion of the Vela supernova 11,500 years ago affected the terrestrial climate[2] coinciding in date with the Biblical flood[53,54]. The sensitivity of the Earth's mesosphere to energetic radiation from nearby supernova explosions also depends on the attenuation of radiation by intervening interstellar gas.

We can calculate the reduction of solar $L\alpha$ flux at Earth due to intervening interstellar $H^o$ associated with a small dense cloud. A window in the molecular hydrogen cross section allows solar $L\alpha$ radiation to penetrate to the upper mesospheric, where the molecular and ozone chemistry of the mesosphere are affected.[55] Fig. 3 shows that a $n\sim10^5$ $cm^{-3}$ 100 K cloud at rest with respect to the Sun would attenuate the solar $L\alpha$ emission by 70%. This reduction would be in addition to the normal factor of two variation in the $L\alpha$ flux between solar cycle minimum and maximum.

An encounter with a dense interstellar cloud would affect the coupling between the terrestrial magnetosphere and solar wind. A high flux of interstellar neutrals at 1 AU would yield a high flux of pick-up ions due to charge exchange with the solar wind. This would change the plasma properties external to the magnetosphere, modify pick-up ion fluxes at Earth, and therefore modify the magnetosphere-solar wind coupling. In turn, this would alter the global electrical circuit (which is function of inflowing cosmic rays, energetic solar particles, the ionospheric electrical properties, and thunderstorm activity[56]). The anomalous cosmic ray population consists of accelerated pick-up ions, and therefore an increase in anomalous cosmic ray fluxes should accompany the increase in pick-up ion densities, depending on details of the acceleration and propagation models. If this is true, the $^{10}$Be ice spikes could also be attributed to an increase in interstellar neutral densities in the inner solar system.

## Conclusions

As the Sun travels through the Milky Way galaxy, the local environment of the Sun changes as the Sun passes through interstellar clouds. Over the past five million years, the Sun has traveled through a region of space between spiral arms, which are defined by the configuration of nearby molecular clouds. Superbubbles formed in the Orion and Scorpius-Centaurus associations have expanded asymmetrically into this interarm region. The nearest portions of the interarm region have been denoted the Local Bubble. Within the past ~$2 \times 10^5$ the Sun emerged from this interarm region and entered the "Local Fluff" cloud complex, which is flowing past the Sun from the direction of the Scorpius-Centaurus Association. The Local Fluff cloud complex is characterized by enhanced interstellar abundances of refractory elements, due to the destruction of dust grains in interstellar shocks apparently associated with the Scorpius-Centaurus superbubble. It appears the Sun entered the low density cloud fragment in which it is presently located within the past few thousand years. Within the next ~2,500 years the cloud boundary inferred to be between the Sun and α Cen should pass over us. For the next million years, nearby gas in the galactic center hemisphere of the sky will sweep past the Sun.

The solar wind shields the Earth from interactions with the low density interstellar cloud now surrounding the solar system. Beyond ~5 au, the densities of the solar wind and inflowing neutral hydrogen are approximately equal. As a result the atmospheres of the outer planets will be more sensitive to changes in the inflowing interstellar matter due to changing interstellar pressure or solar cycle variability. The $^{10}$Be spikes in Antarctic ice cores must be related to structures in the Local Fluff cloud complex, since it is difficult to understand the low velocity dispersion of the Local Fluff cloud complex if it had been recently disrupted by an interstellar shock front. While highly speculative, it is possible that the Local Fluff cloud complex contains small dense cloud wisps that might alter the terrestrial climate, possibly through the reduction of the flux of solar Lα photons to the mesosphere or through variations in the charged particle fluxes that affect the global electrical circuit. This author does not believe that it is a coincidence that the Sun has been immersed for the past few million years in a region of space empty of interstellar clouds, since the development of technological civilizations is more likely to be successful where planetary climates have not been stressed by extreme changes in the interplanetary environments due to "galactic weather".

## Acknowledgments
I would like to thank Carl Heiles for a critical reading of this manuscript, and Chris Wellman for the artwork in Fig. 1.

---

## Figure   Captions

Fig. 1:   The distribution of interstellar matter shown in this figure is compiled from Lucke (1978), Dame et al. (1987), Frisch & York (1983), Welsh et al. (1994), and Frisch (1995).   The yellow arrow is the solar motion, and the light blue arrow is the motion of the interstellar cloud surrounding the solar system, both in LSR coordinates.   The maroon circles are CO molecular clouds.   Other pink and maroon area are regions of diffuse interstellar gas and dust; cloud structure in the lighter pink regions are poorly defined.   The three asterisks are three subgroups of the Scorpius-Centaurus Association.   The three-sided stars in the Geminga Pulsar.   The blue arc towards Orion represents the Orion's Cloak supernova remnant shell.   The other blue arcs are illustrative of superbubble shells from star formation in the Scorpius-Centaurus Association subgroups.   The smallest (i. e. greatest curvature) shell feature represents the Loop I supernova   remnant.

Fig. 2:   The velocities of $Ca^+$ components in 17 nearby stars are plotted in the LSR versus the local interstellar flow velocity frames.   Only one component per star is plotted.   In Fig. 2a the plotted components are selected to be the components with the smallest velocity in the LSR.   These components have average velocities $-5.46 \pm 9.82$ km s$^{-1}$ in the LSR.   In Fig. 2b the plotted components are selected to be those which have the smallest velocities in the flow frame.   They have velocity $-0.054 \pm 2.24$ km s$^{-1}$ in the flow frame.   The LSR vector is calculated based on a peculiar motion of the Sun with respect to the LSR of $-16.5$ km s$^{-1}$ l=53$^o$, b=+25$^o$ is The heliocentric flow   vector $-26.8$ km s$^{-1}$   from the upwind direction l=6.2$^o$, b=+11.7$^o$ is used (Frisch 1995).   Data for stars δ 0, α Hyi, $\tau^3$ Eri, δ Vel, HR4023, β Crt, ι Cen, δ Her, 51 Oph, α Oph, γ Oph, ζ Aql, α Aql, α Gru, γ Aqr, ε Gru are plotted.

Fig. 3:   The disk averaged solar Lα emission profile (solid line) is seen to be attenuated 22% by an intervening neutral cloud with N(H$^o$)=1.5 x 10$^{15}$ cm$^{-2}$ (n~100 cm$^{-3}$), T=7,000 K, V=0 km s$^{-1}$ (dot-dash line), and 70% by a N(H$^o$)=1.5 x 10$^{18}$ cm$^{-2}$ (n~10$^5$ cm$^{-3}$), T=100 K with V=0 km s$^{-1}$ (dashed line).   The vertical scale is arbitrary.